 \documentclass[review, number, sort&compress, 12pt]{elsarticle}
	\usepackage{lineno}

	\usepackage{url}

 \usepackage{graphicx}

\usepackage{amssymb,amsbsy}

\usepackage{textcomp}

\journal{NIM A}

\begin{document}

\begin{frontmatter}



\title{A double-helix neutron detector \\ using micron-size $^{10}$B powder}


\author{Zhehui Wang\corref{cor1}}
 \ead{zwang@lanl.gov}
\author{C. L. Morris\corref{}}%
\author{J. D. Bacon}
\author{M. I. Brockwell}
\author{J. C. Ramsey}
\cortext[cor1]{Corresponding author.}
\address{Los Alamos National Laboratory, Los Alamos, NM 87545}

\begin{abstract}
A double-helix electrode configuration is combined with a $^{10}$B powder coating technique to build large-area (9 in $\times$ 36 in) neutron detectors. The neutron detection efficiency for each of the four prototypes is comparable to a single 2-bar $^3$He drift tube of the same length (36 in).  One unit has been operational continuously for 18 months and the change of efficiency is less than 1\%. An analytic model for pulse heigh spectra is described and the predicted mean film thickness agrees with the experiment to within 30\%. Further detector optimization is possible through film texture, power size, moderator box and gas. The estimated production cost per unit is less than 3k US\$ and the technology is thus suitable for deployment in large numbers. 

\end{abstract}

\begin{keyword}
Double-helix electrode configuration \sep $^{10}$B powder \sep neutron detection efficiency \sep detector lifetime

\end{keyword}

\end{frontmatter}



\newpage
\section{Introduction}
\label{sec:1}

Neutron detectors are basic tools for nuclear science, technology, and applications~\cite{Batchelor:1952,Sayres:1961,Mills:1962,Knoll:2000,McGregor:2003,Athanasiades:2005,Wang:2011,Hoglund:2012,Salvat:2012,Modzel:2014}. Helium-3 ($^3$He) based neutron detectors have been the preferred choices for neutron counting since 1950s~\cite{Batchelor:1952,Sayres:1961,Mills:1962}, when the buildup of tritium stockpile for hydrogen bombs led to a sufficient supply of $^3$He from the tritium beta-decay. The recently recognized $^3$He shortage, partly due to the desire to deploy a large number of neutron counters world-wide to detect concealed special nuclear materials such as plutonium and enriched uranium, has triggered parallel efforts in developing new detector technologies that do not rely on the $^3$He(n, p)$^3$H process for neutron detection. 

Boron-10 ($^{10}$B) based neutron detectors are favorable choices for $^3$He neutron counter replacement for several reasons. $^{10}$B has a thermal neutron capture (absorption) cross section of 3843 barn~\cite{t2:lanl}, 72\% of that of $^3$He. The Q-value of $^{10}$B(n, $\alpha$)$^7$Li reaction is 2.79 MeV. The total energy released during the $^{10}$B neutron capture is only shared exclusively between the two charged particles $\alpha$ (1.78 MeV) and $^7$Li (1.02 MeV) with a likelihood of 6\%. For the other 94\% probability, a 0.48-MeV $\gamma$-ray photon is released, leaving 2.31 MeV to be shared between the $\alpha$ (1.47 MeV) and $^7$Li (0.84 MeV), which is nevertheless sufficient to generate substantial electrical signals in gases. Although $^{10}$B, unlike $^3$He, does form chemical compounds with hydrogen, carbon, nitrogen, halogens, and others, the elemental form of $^{10}$B is non-toxic, chemically inert in the air at the ambient temperature. Highly enriched elemental $^{10}$B ($\sim$ 99 wt\%) can be purchased for several hundred dollars per mol. The benign chemical nature of elemental $^{10}$B and its similarity to $^3$He in terms of neutron capture have motivated us to develop a `drop-in' replacement for $^3$He proportional counters, that is, to achieve similar performances under similar operational and physical constraints.

Although neutron detectors based on the boron-10 trifluoride ($^{10}$BF$_3$) gas was used earlier than $^3$He for neutron detection, the latest efforts in $^{10}$B neutron counters have mostly focused on non-halogen forms of solid-state $^{10}$B compounds. The poor detector performance at high-pressures has limited $^{10}$BF$_3$ gas to a pressure of 0.5 to 1.0 atmospheres in typical neutron counters~\cite{Knoll:2000}. Larger detector volumes and therefore larger footprints are needed to match the efficiency of $^3$He detectors, which typically operate at 2 to 10 atmospheres and have a larger neutron capture cross section as mentioned above. For a large-scale civilian use, the toxicity of $^{10}$BF$_3$ and reactivity in air are not desirable either. 

We describe a large-area (9 in $\times$ 36 in) rectangular $^{10}$B neutron detector for $^3$He counter replacement. Two distinctive features of the $^{10}$B neutron detectors are: 1.) The use of micron-size $^{10}$B powder and a wet-spray coating technique to form the neutron capture thin films on 0.1-in-thick aluminum plates; and 2.) A pair of thin (0.0008$"$ diameter Au-coated tungsten) and thick (0.007" diameter Be-Cu alloy) wires are wound around the $^{10}$B coated aluminum plates to form a `double-helix' electrode configuration, thus the name double-helix neutron detector. 

Below we first descuss the design principles and constraints, followed by some details of the detector design and construction. Next, we describe detector operation, detector performance and compare the experimental results with the predictions using an analytical model. Further detector optimization options, such as film thickness, film texture, neutron moderation using high-density polyethylene etc, are also included. We conclude that the double-helix neutron detector technology is a high-performance, reliable, low-cost solution to the $^3$He neutron-counter shortage problem.

\section{Detector principles and design constraints}
\label{princ:1}

Design of an efficient fission neutron detector using solid forms of $^{10}$B has to take several lengths into account: MeV neutron thermalization length, thermal neutron capture length or the neutron absorption mean free path ($\lambda_a$), $\alpha$ and $^7$Li ranges ($R^i$'s) in the solids of $^{10}$B, and the ranges of $\alpha$ and $^7$Li in gas. We used an existing high-density polyethylene (HDPE) neutron moderator box for $^3$He in this work. It's possible that the efficiencies shown below could be improved further by optimizing the moderator box, but this study is left out here.

The thermal neutron $\lambda_a$ is 19.9 {\textmu}m for the $^{10}$B solid density of  2.17 g/cm$^3$. The charged particle ranges $R^i$'s in $^{10}$B are calculated using SRIM~\cite{SRIM}, as shown in Table.~\ref{Tb:range1}. The smallest $^{7}$Li range ($R^3$) is less than 10\% of the $\lambda_a$. The large differences between $R^i$'s and $\lambda_a$ make a `multi-layer' detector structure universal: an individual layer thickness ($T_0$) is a fraction of $R^i$'s to allow $\alpha$  and $^7$Li to escape from the $^{10}$B layer and generate signals in the gas. The number of layers is a multiple of the product of two ratios, ($\lambda_a/R^i$)($R^i/T_0$), which easily exceeds 10 for efficient neutron detection. A total thickness of  $\lambda_a$ only results in an efficiency of ($1-1/e$ = 63\%) even in the ideal case. A total thickness of 2$\lambda_a$ and 3$\lambda_a$ could result in significantly higher efficiencies of  86\%  and 95\% respectively, at the price of doubling or tripling the number of layers. Therefore the goal of achieving an effective absorption length two to three times $\lambda_a$ is reasonable. 

It is conceivable that the HDPE moderator box could reflect thermal neutrons back and forth through the boron layers a few times, effectively increasing the neutron absorption length without the actual increase in the number of layers. Such `multiple pass effects' deserve further investigation. In 2-bar $^3$He counters, since the thermal neutron absorption mean free path is 3.5 cm, which is less than the diameter of a 2" tube, multiple pass effects are less critical, in particular for neutrons do not enter the detector exactly along the radial directions.

\begin{table}[htdp]
\caption{Maximum ranges ($R^i$) of the charged products from the $^{10}$B(n, $\alpha$)$^{7}$Li neutron capture process in $^{10}$B solid films.}
\begin{center}
{\renewcommand{\arraystretch}{.60}
\renewcommand{\tabcolsep}{0.5 cm}
\begin{tabular}{ccccc}
\hline
Ion & Energy & Probability & Range  & \\
(Label $i$)&($E_0^i$, MeV)&($w^i$)&($R^i$, $\mu$m) & $R^i/\lambda_a$ \\\hline\hline
$\alpha$ (1)& 1.47 & 94\% & 3.35 & 0.17\\
$\alpha$ (2)& 1.78 & 6\% & 4.15 & 0.21\\
${}^{7}$Li (3)& 0.84 & 94\% & 1.78 &0.09\\
${}^7$Li (4) &1.02&6\%& 2.04 & 0.10\\\hline
\end{tabular}}
\end{center}

\label{Tb:range1}
\end{table}%

Since boron is not a semiconductor at room temperature by itself, the alternatives of using semiconductors for $\alpha$ and $^7$Li detection might not be as attractive as they appear. In $^{10}$B doped p-type silicon detectors, for example,  typical doping concentration ranges from 10$^{15}$ to 10$^{18}$ cm$^{-3}$, which are even less than the $^{10}$B concentration in $^{10}$BF$_3$ gas under the 1-bar standard condition. Correspondingly, $\lambda_a$ is at least 2 m, which is impractically large for a semiconductor detector. Yet another alternative is to reduce $\lambda_a$ by $\sim$ a factor of two by operating neutron moderators at  the liquid nitrogen temperature of $77$ K, but the introduction of cryogens could offset the benefits by slowing down neutron to a fraction of the thermal velocity. 

Here, we use aluminum plates as the substrates for $^{10}$B powder coating. The 0.1" thickness of the aluminum plates,  chosen partly for structural steadiness, well exceeds the ranges of $\alpha$ and $^7$Li. Therefore only one of the charged particles ($\alpha$ or $^7$Li, but not simulatenously) is used for electron-ion pair production in gas. The other one is lost to the substrate due to momentum conservation. The need for both high neutron detection efficiency and gamma-ray signal rejection has given rise to several challenges: a.) Large surface area; b.) Film thickness control; c.) Electrical and micro-phonic noise rejection; d.) Background rejection (gammas and cosmic rays); and last but not the least e.) The detector cost.

\section{Detector design and construction}

\begin{figure}[thbp] 
   \centering
   \includegraphics[width=3in]{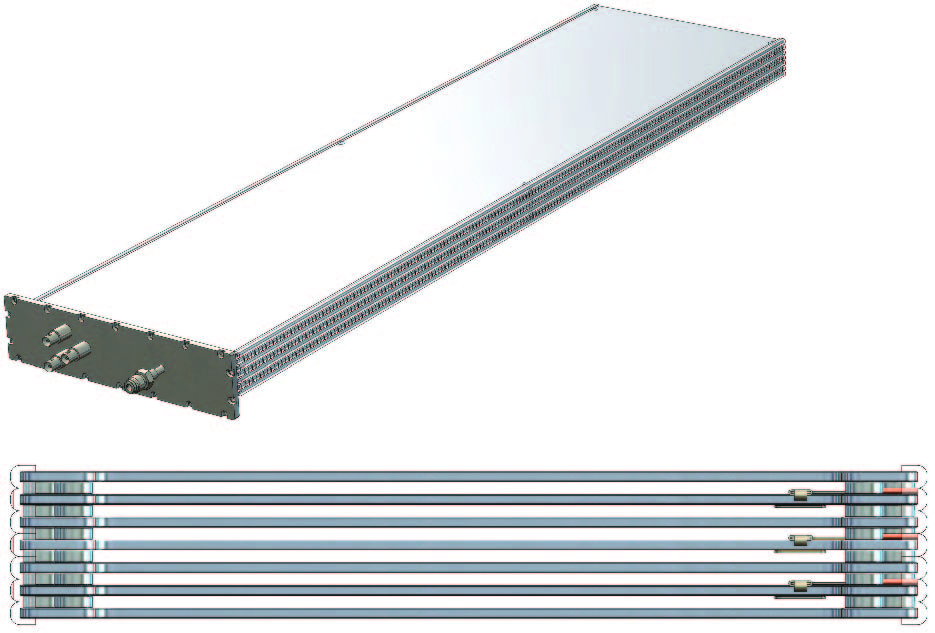} 
   \caption{A twelve-layer $^{10}$B neutron detector core (without the vacuum envelope) viewed from two different angles. The top view shows multiple electrical connectors and the gas port. The bottom view (end-on view, at a larger scale than the top view) shows the insulating acrylic spacers inbetween the layers and the electrical connectors for every other plate.}
   \label{fig:Ramsey}
\end{figure}

The concept of $^{10}$B neutron detector using boron powder was reported previously~\cite{Wang:2011}.  Here we describe a large-area rectangular proportional neutron counter. The same multi-layer detector core is shown in Fig.~\ref{fig:Ramsey} from two different viewing angles. The multi-layer assembly has seven aluminum plates and twelve boron layers. Commercial micron-size $^{10}$B-enriched powder is wet-sprayed onto 36" $\times$ 9" aluminum plates 0.1" thick. Multiple aluminum plates are stacked on top of each other to form a multilayer detector configuration. 

\begin{figure}[thbp] 
   \centering
   \includegraphics[width=4in]{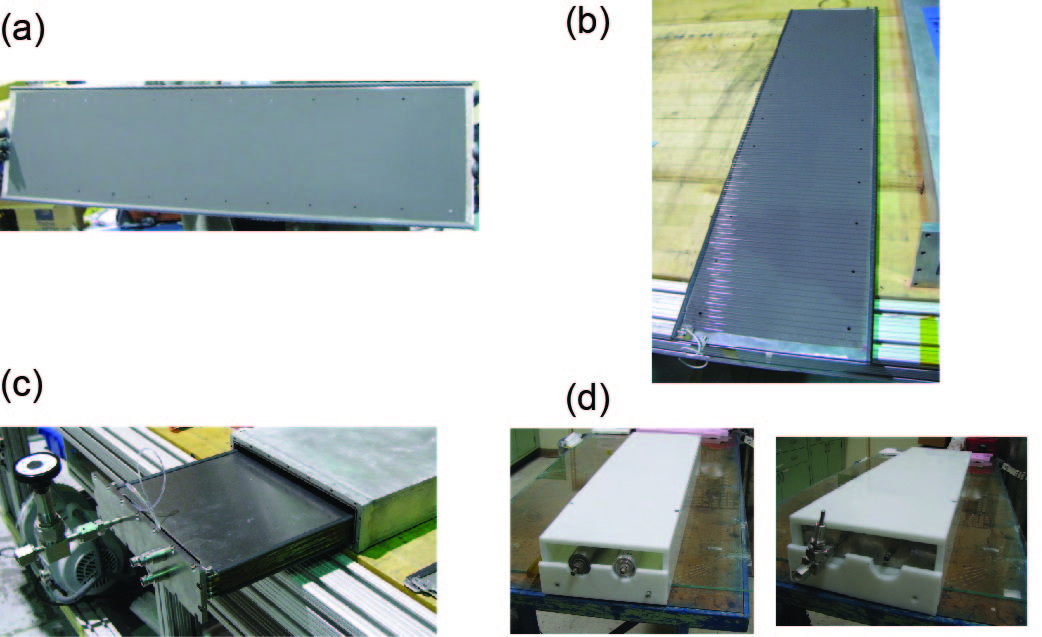} 
   \caption{(a) A $^{10}$B-coated aluminum plate; (b) A coated plate with double-helix winding, also shown in the lower left corner is the wires for biasing and signal transmission; c.) A 12-layer detector core assembled and ready for vacuum sealing and gas fill; d.) A $^{10}$B detector (right) fits inside the same commercial HDPE housing made for two 2-bar $^3$He drift tubes (left).}
   \label{fig:1}
\end{figure}

Both sides of each plate are coated except for the two end plates (the top and bottom), of which only the sides facing inward towards the next plate are coated. The coating thickness is controlled by weighing a predetermined amount of boron to form a liquid solution for wet-spray, 0.3 to 0.5 g per side for the detectors described here. For the surface area of $\sim$1955 cm$^2$, the nominal $^{10}$B film thicknesses are $T_0$ = 0.71  and 1.18 {\textmu}m for 0.3 and 0.5 g of boron respectively. Therefore, the condition $T_0 < R^i$'s is satisfied, according to Table.~\ref{Tb:range1}. A pair of thin (0.0008Ó diameter Au-coated tungsten) and thick (0.007Ó diameter Be-Cu alloy) wires is wound around both sides of every other plate, starting from the next plate from the top, ending at the second plate from the last. Therefore, we always have an odd number (five and seven for the detectors reported here) of plates for our design. The thick wire is the cathode (grounded) and the thin wire is the anode (with a typical positive bias voltage up to 1.4 kV) with a spacing of 4 mm. Anode (cathode) wires from different layers are electrically connected together for a single output (ground). One bar of sealed tetrafluoromethane (CF$_4$) gas is used to stop the ions and generate charge-pairs. The coated plate, double-helix winding, assembly and gas-filled version is shown in Fig.~\ref{fig:1}.

\section{Results and discussion}
\label{sec:DC}
We use commercial pre-amplifiers (ORTEC 142 PC), shaping amplifiers (ORTEC 671), high-voltage supplies (ORTEC 556), and data acquisition (DAQ) system (AMP-TEK MCA model 8000A) for detector characterization. The pre-amp outputs are capacitively coupled to shaping amplifiers by inserting a capacitance of a few nF. The detector capacitance, measured for all twelve layers, is about 270 pF, or about 90 pF per double-helix winding pair that collects signals from four coating layers. The detectors operate in the proportional mode for a typical positive bias voltage between 1.2 kV to 1.4 kV. Therefore, a shaping amplifier only needs to supply a gain up to 10$\times$ with a typical shaping time of a few {\textmu}s. A HDPE-moderated Am/Be neutron source (2.90 mCi when purchased on May 22, 1987) is used for detector characterization. The AmBe currently has a nominal neutron rate of 6.1 kHz (6.4 kHz new). When in contact, a typical count rate exceeds 2000 cps.  

\subsection{Pulse height spectra}
\begin{figure}[thbp] 
   \centering
   \includegraphics[width=3in]{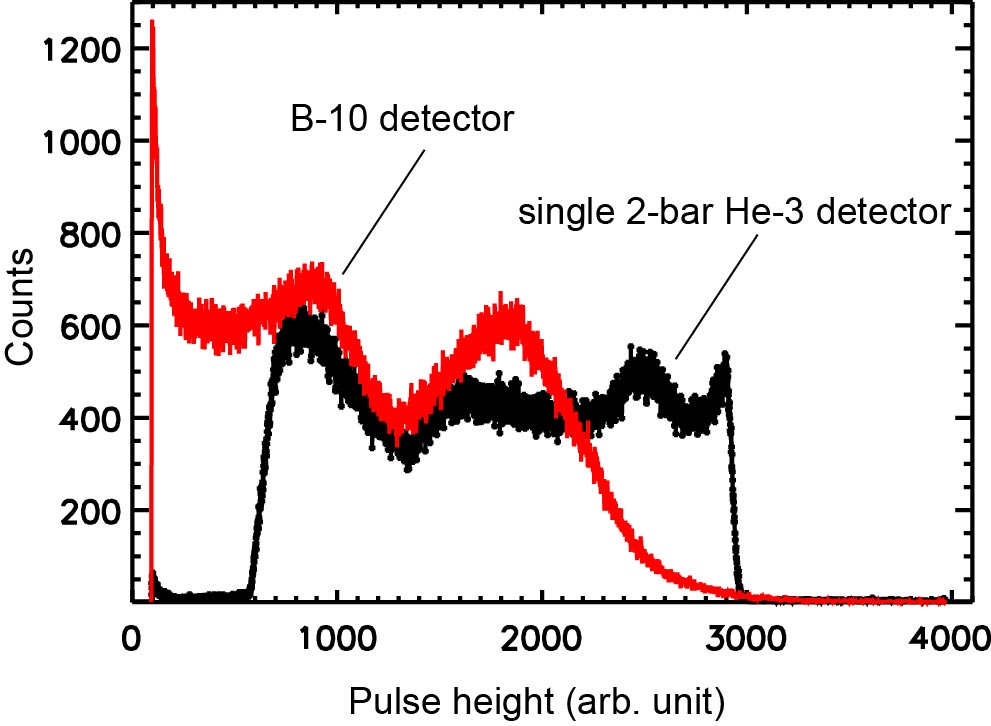} 
   \caption{Pulse height spectrum of a $^{10}$B detector (Red, in the web version of the article) which continuously to the DAQ threshold of 100 mV. Also shown is the PHS of a 2-bar $^3$He detector. The $^{10}$B total neutron count is about 1.2 times of that of $^3$He.}
   \label{fig:2}
\end{figure}

A typical pulse height spectrum (PHS) of a $^{10}$B detector, along with the PHS of a commercial 2-bar 2"-diameter $^3$He detector, is shown in Fig.~\ref{fig:2}. The $^{10}$B PHS extends continuously from the characteristic full energy peaks of $\alpha$ (1.47 MeV) and $^7$Li (0.84 MeV) to the lower energies until the DAQ threshold.  The corresponding 6\% peaks are less prominent. The total number of counts from the $^{10}$B detector is about 120\% of that of the $^3$He. When the neutron count is measured as a function of the source distance, the $^{10}$B detector has an inversely power law dependence with distance with an exponent of -1.46. The $^3$He detector has an exponent of -1.49. The distance-dependences are essentially the same within the measurement error bar.

\subsection{$\gamma$-ray and background sensitivity}
\begin{figure}[thbp] 
   \centering
   \includegraphics[width=3in]{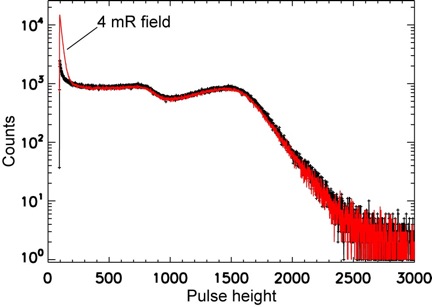} 
   \caption{A 9.8\% neutron efficiency reduction is observed when the detector is placed in a 4-mR field.}
   \label{fig:gammaSens2}
\end{figure}

A preliminary $\gamma$/background-sensitivity study is summarized in Fig.~\ref{fig:gammaSens2}, using 8 pieces of U-238 bricks in contact with the HDPE moderator box. The dose was measured to be about 4 mR. A substantial rise by about 10$\times$ in small amplitude non-neutron signals is observed. The fraction of neutron signal loss is about 9.8\% when raising the discrimination threshold from 100 mV to 200 mV. Additional $\gamma$-sensitivity studies using a more standard setting and sources are being arranged.


\subsection{Stable performance over time}
\begin{figure}[thbp] 
   \centering
   \includegraphics[width=3in]{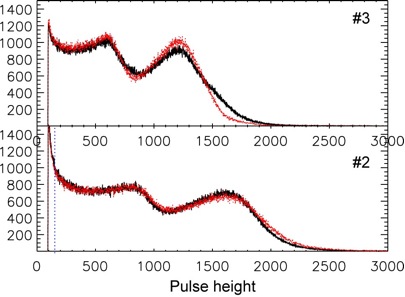} 
   \caption{(Top) Comparison of the PHS of the detector \#3 after 18 months of continuous operation; (Bottom) Comparison of the PHS of the detector \#2 for the same period, during which the detector has mostly been turned off.}
   \label{fig:TimePerform}
\end{figure}

Four prototype detectors were constructed with consistently stable performance. A pair of PHS for two of the detectors are shown in each frame of Fig.~\ref{fig:TimePerform}. Each of the pairs are taken 18 months apart, during which the detector $\#3$ has been running essentially non-stop while \#2 has been mostly off. Both detectors show less than 1\% change in efficiency, which is equivalent to about single 2-bar 2"-diameter 36"-long $^3$He drift-tube detector. The reported efficiencies are cross-calibrated against the commercial $^3$He detector.

\subsection{A theoretical model for PHS}

The energies deposited by $\alpha$ and $^7$Li nuclei in the CF$_4$ gas give rise to PHS. Neglecting the energy straggling, which can be accounted for using a variable gaussian function as shown below, there is a one-to-one correspondence between the energy deposition and the distance ($x$) a nucleus traversing in the $^{10}$B film, $\displaystyle{E^i = E_0^i - \int dx \frac{dE^i}{dx}}$, with $E_0^i$ being the initial kinetic energies of $\alpha$ or $^7$Li, as summarized in Table.~\ref{Tb:range1}.

An observed pulse height spectrum ($dn/dE$) as in Fig.~\ref{fig:2} to Fig.~\ref{fig:TimePerform} is a sum of contributions from $\alpha$ and $^7$Li ions, each with two different initial energies due to the branching ratios,
\begin{equation}
\frac{dn}{dE} = \sum_i w^i \frac{dn^i}{dE'}  \otimes f(E,E'),
\label{eq:sum1}
\end{equation}
in which $w^i$ are listed in Table.~\ref{Tb:range1}. The symbol $\otimes$ stands for the convolution with an instrumental function $f(E,E') = f (E-E')$, which only depends on the differences between two energies in PHS and can be approximated by a normalized gaussian with an pulse-height-dependent width $\sigma_E$~\cite{MW68}, 
\begin{equation}
\sigma_E^2 = \sigma_1^2 + \sigma_0^2\frac{E}{E_0}.
\label{gaus:wd}
\end{equation} 
The first term accounts for thermal, electronics and other energy (pulse height)-independent broadening. The second term, which is energy dependent,  accounts for energy straggling. There is no difficulty to extend the theoretical frame work to more complicated  instrumental functions such as a skewed gaussian. We use Eq.~(\ref{gaus:wd}) here and throughout that captures the essential physics without sacrificing algebraic simplicity. 

\begin{figure}[thbp] 
   \centering
   \includegraphics[width=2.5in]{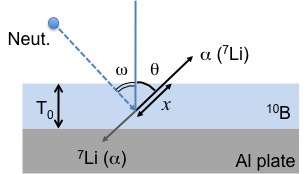} 
   \caption{In a uniform thin film approximation, the neutron absorption efficiency, as well as the resulting PHS, is a function of the neutron incident angle ($\omega$), the distance an ion travels in the film ($x$) and the film thickness ($T_0$).}
   \label{fig:neuteff2}
\end{figure}

The partial spectrum for each species, $dn^i/dE$, is given by
\begin{equation}
\frac{dn^i}{dE} = \frac{dn^i}{dx} (\frac{dE}{dx})^{-1},
\end{equation}
which is a ratio of a geometrical factor $dn^i/dx$ to $dE/dx$, the stopping power of ions in a $^{10}$B film. $dE/dx$ is obtained from SRIM. For a uniform thin film of thickness $T_0$, as illustrated in Fig.~\ref{fig:neuteff2} and an ensemble of neutrons at the same incident angle $\omega$, the average geometrical factor $dn^i/dx$ for each neutron is derived analytically to be
\begin{equation}
\frac{dn^i}{d\tilde{x}} = \frac{-\tilde{R}_x e^{-\tilde{R}_x}+(1-e^{-\tilde{R}_x})}{2\tilde{x}^2},
\label{eq:angular}
\end{equation}
with $\tilde{R}_x= \min(\tilde{x}, \tilde{T}_0)$, the smaller value of the distance an ion travels in the film and the film thickness $T_0$. The factor of $1/2$ is due to the 50\% charge loss to the substrate, the same factor automatically avoids double counting of $\alpha$ and $^7$Li in a neutron capture event. Here $\tilde{x }$, $\tilde{T}_0$ and $\tilde{R}_x$ symbolize the corresponding lengths of $x$, $R_x$ and $T_0$ normalized to $\lambda_a\cos \omega$.

\begin{figure}[thbp] 
   \centering
   \includegraphics[width=3in]{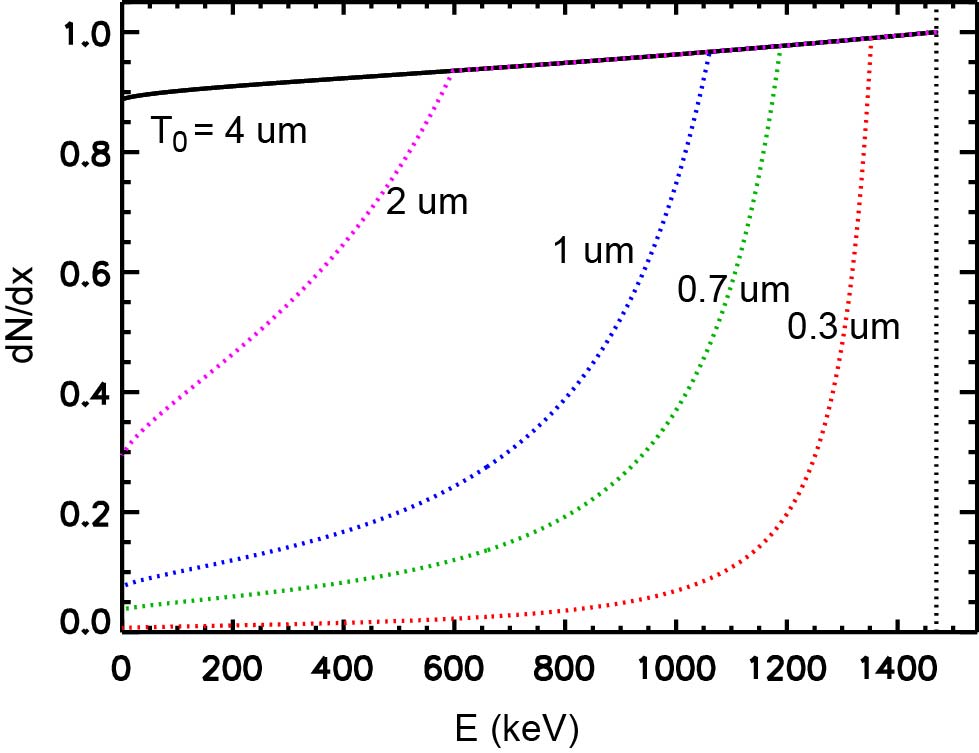} 
   \caption{The geometrical factor of the $^{10}$B film for 1.47 MeV $\alpha$'s, $dn^1/dx$, as a function film thickness ($T_0$). Normal incidence ($\omega = 0$) is assumed for thermal neutrons.}
   \label{fig:thick}
\end{figure}

\begin{figure}[thbp] 
   \centering
   \includegraphics[width=3in]{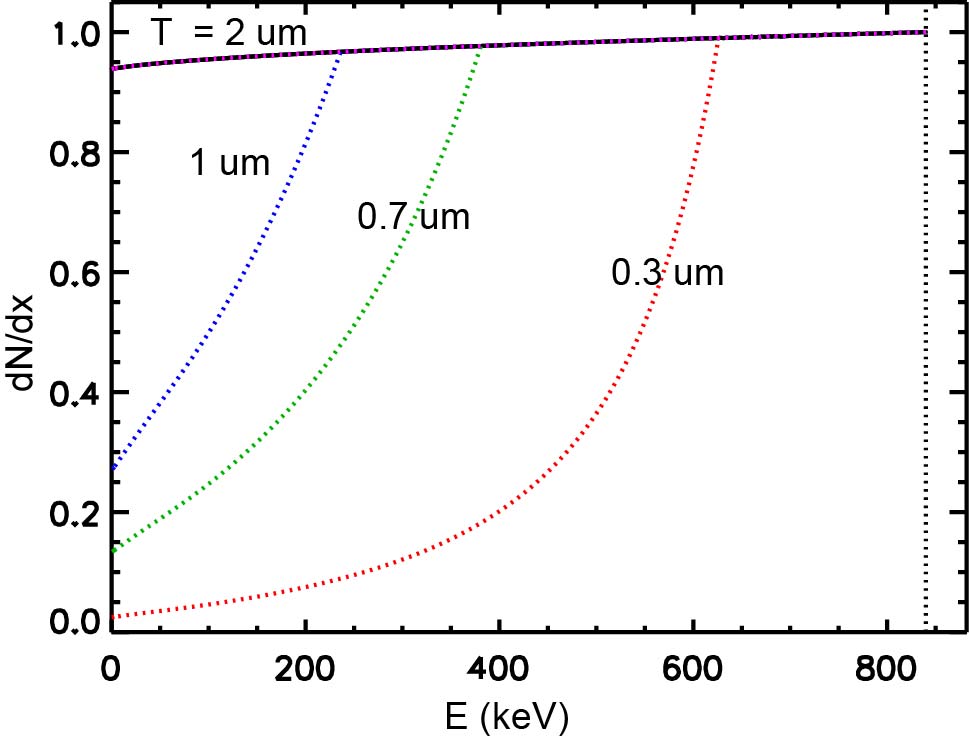} 
   \caption{The geometrical factor of the $^{10}$B film for 0.84 MeV $^7$Li nuclei, $dn^3/dx$, as a function film thickness ($T_0$). Normal incidence ($\omega = 0$) is assumed for thermal neutrons.}
   \label{fig:Li7Th}
\end{figure}

The geometrical factors $dn^i/dx$ as given by Eq.~(\ref{eq:angular}) are plotted for several film thicknesses in Fig.~\ref{fig:thick} for $\alpha$ at 1470 keV and Fig.~\ref{fig:Li7Th} for $^7$Li at 840 keV. The angle of incidence $\omega = 0$. Both $\alpha$ and $^7$Li show a similar dependence on the film thickness. When the film is too thick, $\gtrsim$ 4 {\textmu}m for $\alpha$ and 2 {\textmu}m for $^7$Li, the $dn^i/dx$ distribution does not change, although in practice neutron detection efficiency decreases as the film becomes thicker since neutron captures deep in the film are not recorded. For thinner films, $dn^i/dx$ due to higher energies is the same as in the thick film limit, the lower energy tail gets smaller as the film becomes thinner. It's not surprising that thinner films are better for both higher neutron efficiency and $\gamma$/background discrimination. It's not necessary to go to nanometer thicknesses in practice however, both to avoid an impractically large number of layers and to reduce $\gamma$-sensitivity from the substrate.


\subsection{Comparisons between the theory and experiment}

\begin{figure}[thbp] 
   \centering
   \includegraphics[width=3.5in]{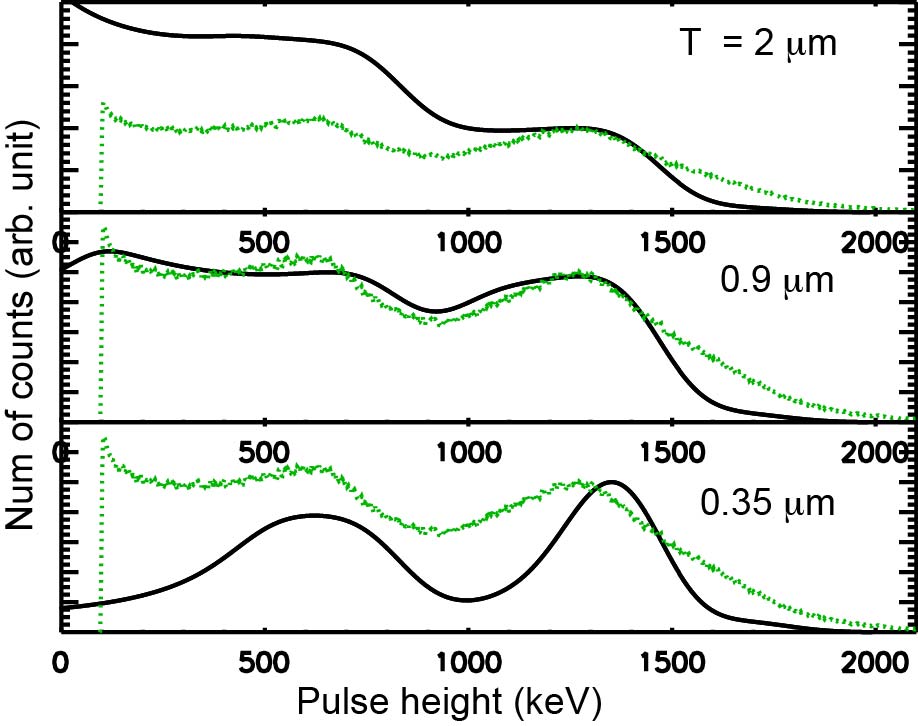} 
   \caption{Theoretically constructed PHS for three different film thickness, 0.35 {\textmu}m, 0.9 {\textmu}m and 2 {\textmu}m. An experimental spectrum is also included for each model. The optimal fit between the experiment and theory is for $T_0$ = 0.9 {\textmu}m.}
   \label{fig:PHSFit3}
\end{figure}

We combine the analytical geometrical factor, Eq.~(\ref{eq:angular}), with SRIM calculations for $dE/dx$ and the variable-width gaussian instrumental function to generate $dn/dE$ as given by Eq.~(\ref{eq:sum1}). The theoretical results for three different film thicknesses, 0.35 {\textmu}m, 0.9 {\textmu}m and 2 {\textmu}m, along with an experimental spectrum, are shown in Fig.~\ref{fig:PHSFit3}. The 0.9 {\textmu}m model best fits the experimental data, although there are apparent disagreements among the details.  The predicted film thickness of 0.9 {\textmu}m is within 30\% of the nominal film thickness of 0.71 {\textmu}m from the experiment. 

Integration of Eq.~(\ref{eq:sum1}) over energy gives the absolute thermal neutron detection efficiency for thin films and loss of efficiency as a function of discrimination threshold. The single-layer thermal neutron detection efficiency for a 0.9-$\mu$m thick $^{10}$B layer and a threshold based on the $\gamma$-sensitivity given in Fig.~\ref{fig:gammaSens2} is
\begin{equation}
\epsilon_1 = \frac{1}{2}\sum_iw^i\epsilon_1^i \sim 86\%.
\end{equation}
Multiple layer effects reduce the average single-layer thermal neutron detection efficiency by 10 to 20\%. Similarly, one obtains the desirable film thickness for a certain target charge collection efficiency, as illustrated in Table.~\ref{Tb:LayerThick1} for at least 90\% charge collection efficiency in each layer. The maximum allowed film thickness, determined by the least energetic $^7$Li at 840 keV, is about 0.4 $\mu$m, or half the values used in the experiment. 

\begin{table}[htdp]
\caption{The maximum film thickness for 90\% charge collection efficiency in a single layer.}
\begin{center}
{\renewcommand{\arraystretch}{.60}
\renewcommand{\tabcolsep}{0.5 cm}
\begin{tabular}{cccc}
\hline
Ion & Energy & Max. $T_0$ & \\
($i$)&($E_0^i$, MeV)&($T_{\max}^i$, {\textmu}m) & $T_{max}^i/R^i$\\\hline\hline
$\alpha$ (1)& 1.47 & 2.0 &60\%\\
$\alpha$ (2)& 1.78 & 3.0 &72\%\\
${}^{7}$Li (3)& 0.84 & 0.4 &22\%\\
${}^7$Li (4) &1.02& 0.6 &29\%\\\hline
\end{tabular}}
\end{center}

\label{Tb:LayerThick1}
\end{table}%

\begin{figure}[thbp] 
   \centering
   \includegraphics[width=3in]{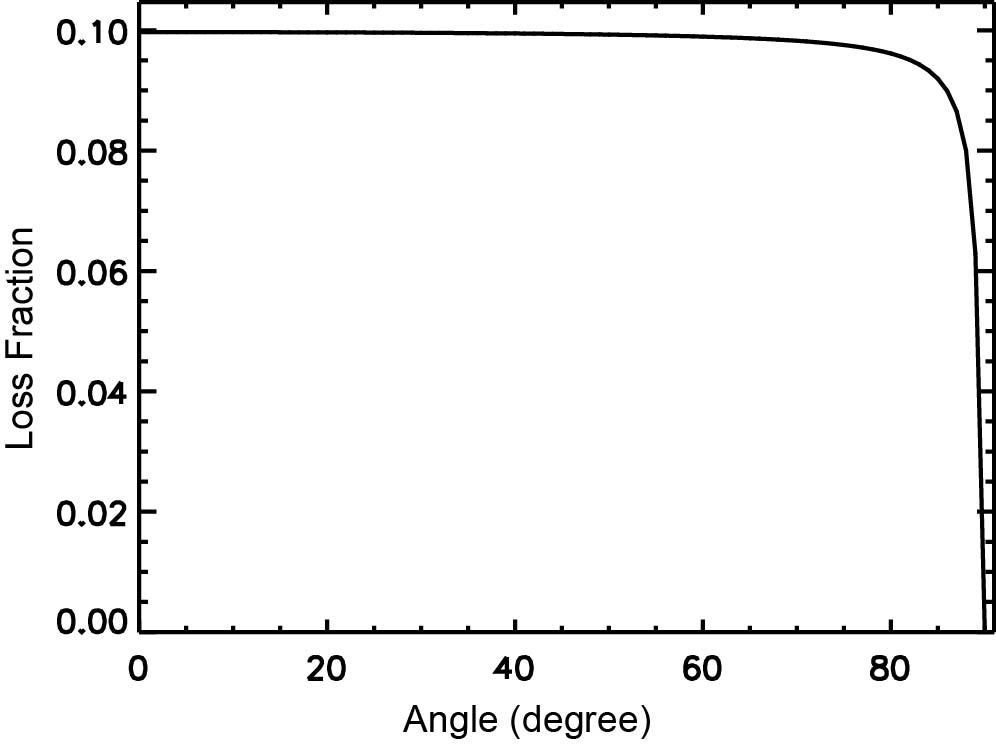} 
   \caption{The reduction of neutron loss as a function of incident angle. The improvement in neutron efficiency is insignificant for the angles less than 85 degrees. The result is for a $^{10}$B film 0.7-$\mu$m thick and thermal neutrons.}
   \label{fig:neutangle2}
\end{figure}

It should be mentioned that thus far our analysis has assumed $\omega = 0$, or normal neutron incidence. This is justified by Fig.~\ref{fig:neutangle2}, which shows that the potential benefits of larger angles of incidence with a fixed discrimination threshold. The benefit is insignificant until the incident angles exceed 85 degrees, which account for less than a few percent of the total neutron population expected. 



\subsection{Powder size and film texture}
\begin{figure}[thbp] 
   \centering
   \includegraphics[width=2in]{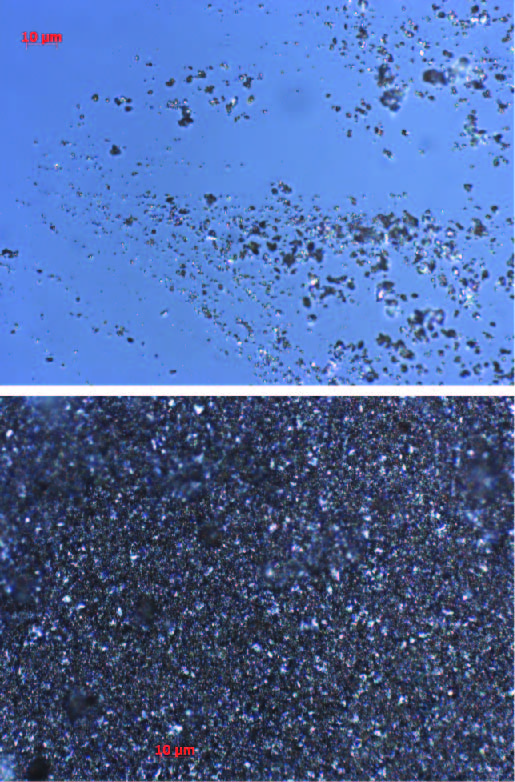} 
   \caption{(Top) $^{10}$B powder under an optical microscope, showing a large dispersion of powder sizes (up to 10 {\textmu}m according to the supplier) and powder agglomeration. (Bottom) Powder coating on a detector substrate shows very rough (on the $\mu$m)-scale film texture)}
   \label{fig:powder}
\end{figure}

A microscopic image of the powder used and of a powder coating on a detector substrate are shown in Fig.~\ref{fig:powder}. The large dispersion of the powder sizes, ranging from sub-{\textmu}m to about 10 {\textmu}m, is consistent with the supplier specifications. The texture of the powder coating on the detector surface is far away from being a text-book smooth surface on the length scale that is comparable to the ion ranges in $^{10}$B or the nominal film thicknesses, as assumed above. The implications of such complex texture for neutron detection shall be left for further studies. We only point out here that a.) The ideal thin film, a thin slab with a uniform thickness $T_0$, is not the geometry to achieve the largest surface area; and b.) How artificial surface textures could be used to significantly increase the surface area, thus allowing more $^{10}$B on a fixed surface and maximizing the efficiency without resorting to too many surfaces.

One may divide a large flat surface into small tiles of squares with an area $A_0$. The total $^{10}$B mass of each square is proportional to the volume $A_0T_0$.  Using the same footprint $A_0$ and the same amount of $^{10}$B mass (volume) to form a hemisphere, the surface area of the hemisphere is $\displaystyle{\frac{\pi}{2} A_0}$, which is 1.57 times that of the square tiles. Meanwhile, the area $A_0$ is related to the thickness $T_0$ as $\displaystyle{T_0=\frac{\pi}{12} \sqrt{A_0}}$  in order that the volume of each hemisphere matches the square tile. If one uses regular tetrahedrons as the tiles, the total surface area of the three sides for each tetrahedron is obviously $3A_0$. Correspondingly, the area $A_0$ is related to the thickness $T_0$ as $\displaystyle{T_0 =\frac{2^{3/2}}{3^{7/4}}\sqrt{A_0} \sim 0.41 \sqrt{A_0}}$  to match the volume. Therefore, either using the hemisphere or the regular tetrahedron can obtain larger surface area than the planar thin film for the same amount of boron and surface (footprint) area.

\begin{figure}[thbp] 
   \centering
   \includegraphics[width=3in]{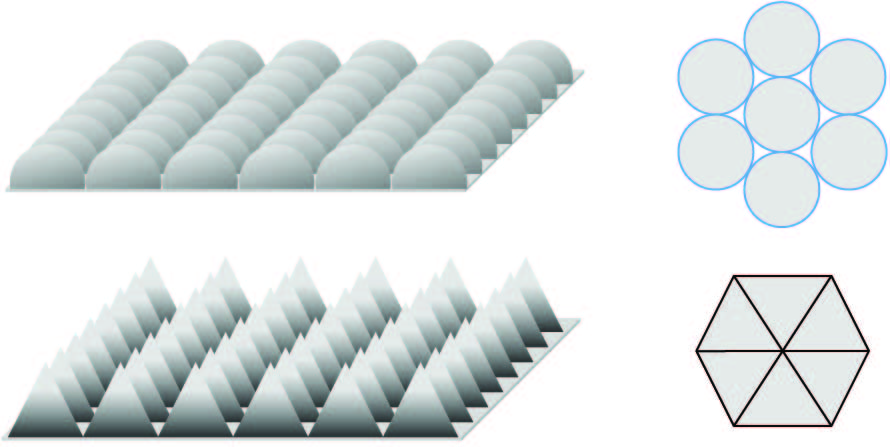} 
   \caption{(Top) `Three-D' rendition of a hemisphere tiling [left] and the corresponding footprint [right]. (Bottom) `Three-D' rendition of a regular tetrahedron tiling [left] and the corresponding footprint [right]. By replacing conventional slab films with three-dimensional micron-size objects as tiles, 1.5 to 3 times as much surface area as a slab film can be achieved. This approach if can be demonstrated experimentally, could relax the reliance on using too large number of surfaces for high neutron detection efficiency.}
   \label{fig:texture2}
\end{figure}

\section{Conclusions}

We have experimentally demonstrated a large-area (9 in $\times$ 36 in) neutron detector that combines a double-helix electrode configuration with a $^{10}$B powder coating technique. The neutron detection efficiency for each of the four prototypes is comparable to a single 2-bar $^3$He drift tube of the same length (36 in).  One unit has been operational continuously for 18 months and the change of efficiency is less than 1\%. Another unit, which was turned off for the same period, does not show any significant change in the efficiency either. A 4-mR $\gamma$/background test indicates the need for raising the pulse height discrimination threshold, leading to a reduction of efficiency by about 10\%. An analytic model for pulse heigh spectra is described and the predicted mean film thickness agrees with the experiment to within 30\%. Further detector optimization is possible through film thickness, film texture, power size, moderator box and gas. The cost of each detector, including raw material, fabrication, and labor, is expected to be less than US\$ 3k. 

{\bf Acknowledgments} We thank Mr. Michael Everhart-Erickson, Ms. Laura Barber and Ms. Erica Sullivan for their efforts in commercialization of the $^{10}$B-based neutron detector technology. This work was supported in part by a CRADA agreement with TSA systems, Longmont, CO, USA.

\bibliographystyle{elsarticle-num}
\bibliography{DHB10v2}

\providecommand{\noopsort}[1]{}\providecommand{\singleletter}[1]{#1}%
\begin{thebibliography}{10}
\expandafter\ifx\csname url\endcsname\relax
  \def\url#1{\texttt{#1}}\fi
\expandafter\ifx\csname urlprefix\endcsname\relax\def\urlprefix{URL }\fi
\expandafter\ifx\csname href\endcsname\relax
  \def\href#1#2{#2} \def\path#1{#1}\fi

\bibitem{Batchelor:1952}
R.~Batchelor, R.~Aves, T.~H.~R. Skyrme, Rev.\ Sci.\ Instr. 26 (1955) 1037.

\bibitem{Sayres:1961}
A.~R. Sayres, K.~W. Jones, C.~S. Wu, Phys.\ Rev. 122 (1961) 1853.

\bibitem{Mills:1962}
W.~R. Mills, R.~L. Caldwell, I.~L. Morgan, Rev.\ Sci.\ Instr. 33 (1962) 866.

\bibitem{Knoll:2000}
G.~F. Knoll, Radiation Detection and Measurement, 3rd Edition, John Wiley \&
  Sons, 2000.

\bibitem{McGregor:2003}
D.~S. McGregor, et~al., Nucl.\ Intrum.\ Meth.\ A 500 (2003) 272--308.

\bibitem{Athanasiades:2005}
A.~Athanasiades, N.~N. Shehad, C.~S. Martin, et~al., IEEE Nucl. Sci. Symp.
  Conf. Rec. 2 (2005) 623--627.

\bibitem{Wang:2011}
Z.~Wang, C.~L. Morris, Nucl.\ Intrum.\ Meth.\ A 652 (2011) 323--325.

\bibitem{Hoglund:2012}
C.~H{\"o}glund, J.~Birch, K.~Andersen, et~al., J.\ Appl.\ Phys. 111 (2012)
  104908.

\bibitem{Salvat:2012}
D.~J. Dalvat, C.~L. Morris, Z.~Wang, et~al., Nucl.\ Intrum.\ Meth.\ A 691
  (2012) 109--112.

\bibitem{Modzel:2014}
G.~Modzel, M.~Henske, A.~Houben, et~al., Nucl.\ Intrum.\ Meth.\ A 743 (2014)
  90--95.

\bibitem{t2:lanl}
R.~MacFarlene, I.~A.~C.~Kahler, P.~Moller, M.~W. Paris,
  \url{http://t2.lanl.gov/data/endf/index.html} (2012).

\bibitem{SRIM}
J.~F. Ziegler, \url{http://www.srim.org/} (2012).

\bibitem{MW68}
G.~L. Morgan, R.~L. Walter, Phys. Rev. 168 (1968) 1114.

\end{thebibliography}







\end{document}